# Optical Coupling of Mie Particles Adsorbed on a Whispering Gallery Resonator

[o][*]**Frank Vollmer,** [*]**Hai-Cang Ren,** [#]**Stephen Arnold,** [*]**Albert Libchaber**

[*] *Center for Studies in Physics and Biology, The Rockefeller University, 1230 York Ave, New York, NY 10021*

[#]*Microparticle Photophysics Laboratory, Polytechnic University, 6 Metrotech Center, Brooklyn, NY 11201*

[o]corresponding author; email: vollmef@rockefeller.edu

## Abstract

We present an experimental observation on particles adsorbed to a whispering gallery mode (WGM) optical resonator produced from a glass sphere 100 µm in radius. For spherical adsorbates of radius a 100 nm (meso-optic, 2 a/ ~1) we find a striking nonlinear dependence of the induced wavelength shift on the adsorbed particle density, whereas for Rayleigh particles with dimensions much smaller than the infrared wavelength the dependence is linear with particle density. An explanation of the anomalous effect based on optical coupling between meso-optic adsorbates is suggested.

Ever since the pioneering work on high Q whispering gallery resonances in spherical micrometer sized resonators by Ashkin and Didzkic [1] there have been a plethora of articles [2] which have extended the measured "ultimate Q" [3] to ~$10^{10}$. The interest in such resonators has been fueled by a diversity of areas from studies of strong coupling in quantum electrodynamics [4] to ultra-sensitive bio-sensing [5]. The latter interest, which has produced a record sensitivity, depends on the shift in resonance frequency due to the perturbation by adsorbed nanoparticles (i.e. protein molecules, DNA, etc.).

The recent experiments on DNA hybridization [6] and specific protein interactions [5] deal with adsorbates ~10 nm in size for which a first order perturbation theory of Rayleigh scattering has successfully been applied [7, 8]. Herein we present experiments on adsorbates in transition toward Mie sizes (meso-optic, size ~100 nm) and find an anomalous enhancement over the first order theory. In addition, our results have a fundamentally different form. Although the Rayleigh theory predicts a resonant wavelength shift in proportion to surface density, with meso-optic adsorbates the effect has a predominant density squared component. It appears that the new phenomenon is due to resonantly enhanced coupling between the adsorbed particles.

In what follows we briefly describe our experimental approach, present the anomalous results on meso-optical particles and outline a theoretical approach based on the multiple scattering of the WGM by adsorbed particles.

In our experiments we use a microsphere cavity as an ideal optical resonator for WGMs. Smaller particles can be bound to such a cavity. Such adsorbed particles perturb the WGM by interaction with the associated evanescent field. We use highly polarizable polystyrene spheres (called nanospheres) with a refractive index of n = 1.59 of varying size to perturb the WGM (Fig.1).

We excite WGMs inside the silica microsphere cavity (n = 1.46) of typical radius R = 110 µm by evanescent coupling to a single mode fiber [9] which is eroded into its core by etching with hydrofluoric acid. [10]. The microspheres are made by melting the tip of a single mode fiber in a butane/nitrous oxide flame [11]. The fabricated sphere is mounted on a xyz stage and positioned in mechanical contact with the eroded fiber core, to enable evanescent coupling. Coherent light is transmitted through the single mode fiber from a current tunable distributed feedback laser operating at a nominal wavelength = 1312 nm. The fiber-transmitted intensity is detected by a photodetector at the other fiber end. We identify the resonance wavelength from the minima of a Lorentzian-shaped dip in the transmission spectrum [12]. This microsphere-fiber system is operated in an aqueous environment (salted phosphate buffer PBS, pH 7.4) with a Q ~2 x $10^6$. The resonance wavelength is determined with a precision ~1/$50^{th}$ of the linewidth from a parabolic minimum fit of the resonance dip. A suspension of nanospheres of a given size is then injected into the liquid filled sample cell whereupon the nanospheres diffuse to and stably adsorb on the surface of the microsphere cavity. The resonant wavelength vs. nanosphere surface density is continually monitored from the point of injection.

We use fluorescent, carboxylated polystyrene nanospheres to study the perturbative effect. The sample cell is examined with a standard fluorescence microscope equipped with a Xenon lamp and fluorescent filter set. The surface density of adsorbed yellow-green fluorescent nanospheres (Molecular Probes) is determined from an image taken with a cooled CCD camera. We plot the recorded wavelength shift versus the nanosphere surface density.

For nanospheres of meso-optical size ~100 nm radius we observe an anomalous shift of the resonant line (Fig. 2). The wavelength shift is much larger than expected from our first order perturbation theory developed for Rayleigh particles. Furthermore, we observe a pronounced nonlinear dependence of the resonant line shift with nanosphere surface density $\sigma$. This effect is not seen for nanospheres of radii ~50 nm or smaller where the Rayleigh theory [7] applies. The mean field approximation to multiple scattering of nanospheres gives rise to the linear scattering, which is represented in Fig.2 by a solid line for the Rayleigh limit and by the dashed line when the evanescent field penetration depth is taken into account.

For the larger nanospheres the theoretical explanation of the observed effect is challenging because of its departure from the Rayleigh regime and the relatively large polarizability of the adsorbates. A model for this strong nonlinear dependence of the resonant wavelength shift on the density of meso-optic adsorbates will be constructed below.

The electromagnetic field induced in the cavity and the solution by the eroded optical fiber is determined by the macroscopic Maxwell equations with an inhomogeneous dielectric function,

$$\varepsilon = \varepsilon_0 + \delta\varepsilon \tag{1}$$

with $\varepsilon_0$ accounting for the background (cavity and solution) and $\delta\varepsilon$ accounting for the surface adsorbed nanospheres. We have $\varepsilon_0 = \varepsilon_{cav.}$ for $r < R$ and $\varepsilon_0 = \varepsilon_{sol.}$ for $r > R$ with $R$ the radius of the cavity. The dielectric excess of the adsorbed nanospheres is given by

$$\delta\varepsilon(\mathbf{r}) = \Delta\varepsilon \sum_{i=1}^{N} \theta(a - |\mathbf{r} - \mathbf{r}_i|) \tag{2}$$

where $\Delta\varepsilon = \varepsilon_{particle} - \varepsilon_{sol}$, the difference between the dielectric constant of the nanospheres and that of the solution, $\mathbf{r}_i$ is the location of i-th particle, a is the radius of the spherical particle and $N$ ($=4\pi R^2 \rho$) their total number. For a harmonic time dependence, $e^{-i\omega t}$, the linear response of the electric field $\mathbf{E}$ to the source current of the fiber, $\mathbf{J}$, reads

$$E_a(\mathbf{r}) = -i\omega \int d\mathbf{r}' G_{ab}(\mathbf{r},\mathbf{r}'|\omega) J_b(\mathbf{r}'), \tag{3}$$

where $G_{ab}(\mathbf{r},\mathbf{r}|\omega)$ is a Green function tensor component, repeated indices are summed, and the angular frequency $\omega = \dfrac{2\pi}{\sqrt{\varepsilon_{sol.}}\lambda}$ with $\lambda$ the wavelength in the solution. A resonance corresponds to a pole of the Green function in the complex $\omega$ plane, and is labeled by its type, TE (transverse electric) or TM (transverse magnetic), the angular momentum quantum numbers $(l,m)$, and a radial quantum number $n$. In the absence of the adsorbates, $\delta\varepsilon = 0$, the Green function $G_{ab}(\mathbf{r},\mathbf{r}|\omega) = G_{ab}^{(0)}(\mathbf{r},\mathbf{r}|\omega)$ and is explicitly known. The WGMs refer to those narrow resonances of $l \sim \dfrac{2\pi R}{\lambda}$, which have a concentrated profile about the surface of the cavity.

For $\delta\varepsilon \neq 0$, we are interested in the ensemble average of $G_{ab}(\mathbf{r},\mathbf{r}|\omega)$ over random distributions of the adsorbed nanospheres, which is denoted by $G(\mathbf{r},\mathbf{r}'|\omega)$ and satisfies the integro-differential equation [13]

$$\left[\omega_0^2 - \nabla\times(\nabla\times)\right]_{ac} G_{cb}(\mathbf{r},\mathbf{r}'|\omega) + \omega^2 \int d\mathbf{r}_1\, \varepsilon_{ac}(\mathbf{r},\mathbf{r}_1|\omega)\, G_{cb}(\mathbf{r}_1,\mathbf{r}'|\omega) = \delta_{ab}\delta^3(\mathbf{r}-\mathbf{r}'),$$

(4)

where the differential operators inside the bracket act on the first coordinate of the Green function and the effective dielectric function $\varepsilon_{ab}(\mathbf{r},\mathbf{r}'|\omega)$ accounts for the ensemble average of all multiple scattering processes. A resonant mode corresponds to a zero mode of the integro-differential operator on the left hand side of (4), which is analogous to an eigenstate of the hamiltonian operator of a spin one particle in a non-local external potential $-\omega^2\varepsilon_{ab}(\mathbf{r},\mathbf{r}'|\omega)$. The potential can be expanded according to the ascending powers of $\varepsilon$ and , with each term represented by a Feynman-like diagram, one of which is displayed in the inset of Fig.3. The power of $\varepsilon$ represents the order of the Born approximation. The power expansion in nanosphere surface density takes into account the mean field approximation and ~~fluctuation~~ spatial randomness of the adsorbates.

We suspect that the pairing effect depicted in Fig.3 accounts for the coupling between two nanospheres. Mutual scattering of WGMs by the nanospheres produces the major contribution to the observed nonlinearity and we have estimated its magnitude within the framework of the Born approximation. The leading order diagram of the pairing potential is shown in the inset of Fig.3, where a solid line is associated with a free Green function, a filled circle is associated with a vertex function $-\omega^2 \delta\varepsilon(\mathbf{r})$ and an open circle generates a connected average among the vertex functions linked to it through the dashed lines. This diagram is proportional to $\sigma^2 \varepsilon^3$. We can write the resonant line shift as

$$\frac{\ }{\ } = \frac{\ }{\ }\bigg|_{linear} \left(1 + C\ a^2\right), \tag{5}$$

with

$$\frac{\ }{\ }\bigg|_{linear} = \frac{8\ a^3}{(\ _{cav.} -\ _{sol.})R}\, f\!\left(\frac{8\ a}{\ _{sol.}}\right) \sqrt{\frac{\ _{cav}}{\ _{sol.}} - 1} \tag{6}$$

where the form factor $f(z) = \frac{1}{z}(1 + e^{-z}) - \frac{2}{z^3}(1 - e^{-z})$ takes care of the damping of the evanescent field outside the cavity. Notably for Rayleigh particles this expression reduces identically to the result given by [7].

We found that the coefficients C can be approximated by a function of $\frac{a}{\lambda}$ only and the line-shift is inversely proportional to *R*. This is consistent with the observed scaling behavior of the line-shift with the cavity radius. The uncertainties behind this estimate stem from the lack of a clear-cut separation between WGMs and non-resonance modes. From our data measured for nansopheres of radius a = 105 nm at two different cavity sizes (R = 110 µm and 200 µm) we determine the coupling coefficient C ≈ 2500. A robust calculation of the nonlinearity is under way, which will involve pairing effect as well as weak localization.

## Acknowledgment

Frank Vollmer would like to thank the Boehringer Ingelheim Fonds for his PhD fellowship. Hai-cang Ren would like to thank Z. B. Su for an interesting communication. The work of Hai-cang Ren is supported in part by US Department of Energy under the contract no. DE-FG02-91ER40651-TASKB. Stephen Arnold was supported by a grant from the National Science Foundation grant (BES-0119273).

**Figure legends**

**Figure 1: A** – The silica microsphere cavity is evanescently coupled to an etch-eroded single mode optical fiber operated at = 1.3 µm wavelength. The wavelength is scanned by a current tunable laser source. Cavity resonances are detected as Lorentzian-shaped dips in the transmission spectrum recorded by the photodetector at the fiber end. **B -** The microsphere-fiber contact is located in a sample cell filled with a dilute, aqueous suspension of nanospheres. Nanosphere particles adsorbing to the microsphere cavity cause a red-shift of the resonant line. The surface density of the fluorescent particles is determined from an image acquired with a fluorescence microscope. We relate the surface density with the recorded wavelength shift.

**Figure 2:** Anomalous resonance wavelength shift produced by meso-optical polystyrene nanospheres of a = 105 nm radius. This nonlinear effect has a quadratic dependence on the surface density. The data (dots) is also plotted for nanospheres of a = 50 nm radius which follow the linear mean field theory for this Rayleigh limit. The graph shows the predictions of this theory for both nansophere sizes (solid lines). Furthermore the Rayleigh theory is corrected for the evanescent field penetration depth (dashed lines). The onset of the nonlinear effect for nanospheres larger than 50 nm in radius marks the transition between the Rayleigh and Mie regime.

**Figure 3:** Multiple scattering theory explains the nonlinear effect as a pairwise coupling between surface adsorbed nanospheres. The equatorial microsphere cavity resonance

(WGM, quantum numbers n and l) excites a nearby bound nanosphere. This first nanosphere then interacts with a second nanosphere via scattered light ($n_1 l_1$ and $n_1' l_1'$) causing a nonlinear perturbation of the resonance wavelength. The inset shows the Feynman-like diagram explaining the pairing effect, which is quadratic in nanosphere surface density. The microsphere is of typical radius R  100 µm, the nanospheres causing the anomalous wavelength shift are of typical radius a  100 nm.

**Figure 1**

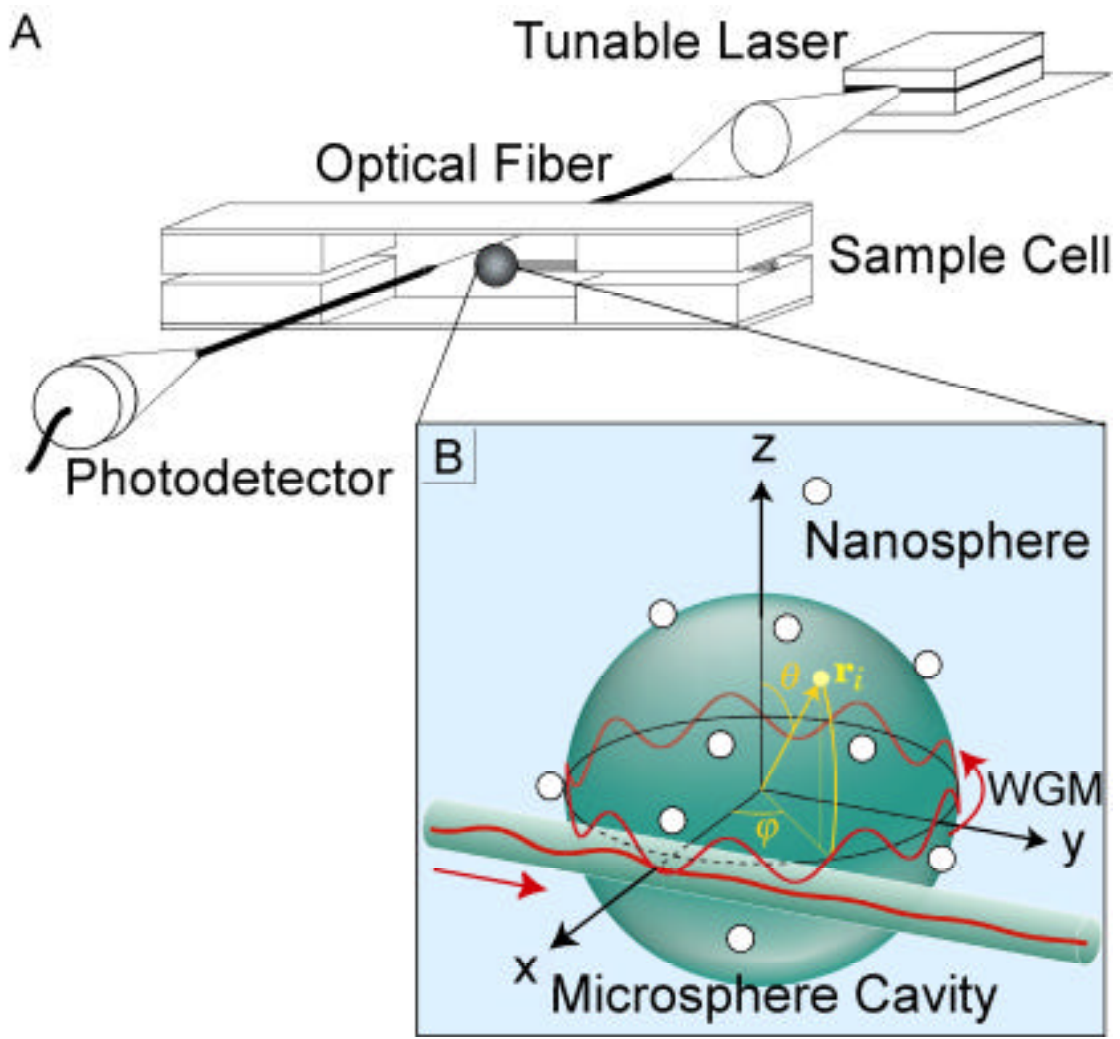

**Figure 2**

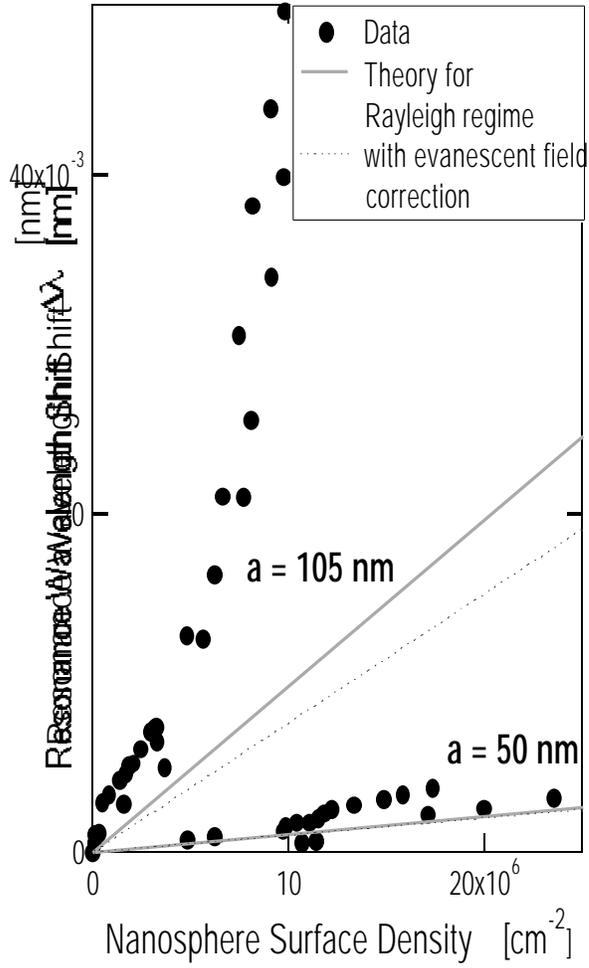

**Figure 3**

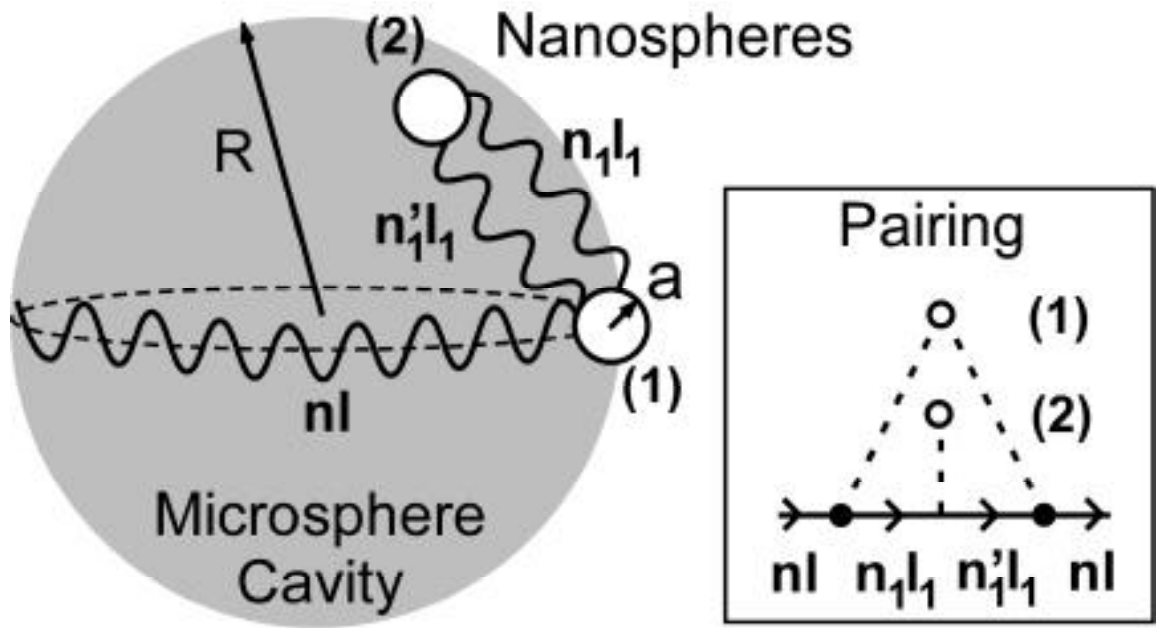